\documentclass[prb,twocolumn,showpacs,preprintnumbers,floats]{revtex4-1}%
\usepackage{graphicx}
\usepackage{amsmath}
\usepackage{epstopdf}
\usepackage{bbold}
\usepackage{amsbsy}
\usepackage{times}
\usepackage{color}
\usepackage{hyperref}
\usepackage{dcolumn}
\usepackage{bm}%
\usepackage{amsfonts}%
\usepackage{amssymb}
\providecommand{\U}[1]{\protect\rule{.1in}{.1in}}
\def\red{\color{red}}

\graphicspath{{./figures/}}

\begin{document}
\title{Detecting sign-changing superconducting gap in LiFeAs using quasiparticle
interference }
\author{D. Altenfeld$^{1}$}
\author{P. J. Hirschfeld$^{2}$}
\author{I.I. Mazin$^{3}$}
\author{I. Eremin$^{1}$}
\affiliation{$^{1}$ Institut f\"ur Theoretische Physik III, Ruhr-Universit\"at Bochum,
D-44801 Bochum, Germany}
\affiliation{$^{2}$Department of Physics, University of Florida, Gainesville, Florida
32611, USA}
\affiliation{$^{3}$Code 6393, Naval Research Laboratory, Washington, DC 20375, USA }
\date{\today}

\begin{abstract}
Using a realistic ten-orbital tight-binding model Hamiltonian fitted to the
angle-resolved photoemission (ARPES) data on LiFeAs, we analyze the
temperature, frequency, and momentum dependencies of quasiparticle
interference (QPI) to identify gap sign changes in a qualitative way,
following our original proposal [Phys. Rev. B 92, 184513 (2015)]. We show that
all features present for the simple two-band model for the sign-changing
$s_{+-}$-wave superconducting gap employed previously are still present in the
realistic tight-binding approximation and gap values observed experimentally.
We discuss various superconducting gap structures proposed for LiFeAs, and
identify various features of these superconducting gaps functions in the
quasiparticle interference patterns.  On the other hand, we show that it will be difficult to identify the more complicated possible  sign structures of the hole pocket gaps in LiFeAs, due to the smallness of the pockets and the near proximity of two of the gap energies.
\end{abstract}

\maketitle

\section{Introduction}

In the iron-based superconductors, the origin of the Cooper-pairing and the
overall phase structure of the superconducting order parameter is still under
debate. Nevertheless, taking into account the Fermi surface topology of iron
based superconductors, there is a general consensus that the superconducting
state in most of these materials belongs to the $A_{1g}$ symmetry
representation as long as both the electron and the hole Fermi surface pockets
are present in the system
\cite{doi:10.1146/annurev-conmatphys-020911-125055,PhysRevLett.107.147002}. At
the same time, due to the multiplicity of the Fermi surface sheets and
multiorbital character of the electron states near the Fermi level, the actual
phase structure of the superconducting gap distributed between various Fermi
surface and orbitals is less clear. While orbital fluctuations together with
the electron-phonon interaction favor the conventional \textquotedblleft%
$s_{++}$" wave superconducting gap without internal phase shifts
\cite{PhysRevLett.104.157001}, spin fluctuations and enhanced interband
repulsion originating from the proximity to the antiferromagnetic phase favor
the so-called $s_{+-}$-wave structure in which the sign of the order parameter
changes between the hole and electron
bands.\cite{PhysRevLett.101.057003,PhysRevLett.101.087004}. Furthermore, the
situation can be even richer if the magnetic fluctuations are not so strong as
it is the case in LiFeAs\cite{Wang2013,Yin2014,Ahn2014}, or when fluctuations
at more than one wavevector compete, as in FeSe\cite{Glasbrenner15,Sprau17}.
In particular, it was argued that in FeSe the electronic renormalization
effects for different orbitals result in  orbital selectivity for the
Cooper-pairing\cite{Sprau17,Kreisel17}, which in turn also affects the overall
magnitude and the phase structure of the superconducting gap. In LiFeAs, orbital selective renormalization was also recently
observed\cite{Kreisel17,PhysRevB.94.201109}. Flavors of the $s_{+-}$ pairing
symmetry where the phase of the order parameter is additionally changing
between different hole or electron Fermi pockets  have been proposed for this or other Fe-based superconductors\cite{Kontani2014,Yin2014,Ahn2014,Maiti2013,Khodas2012}. 

One rapidly developing technique to determine the phase structure of the order
parameter makes use of quasiparticle interference as measured by Fourier
transform scanning tunneling microscopy (FT-STM). This probe measures the
wavelengths of Friedel oscillations caused by disorder present in a metallic
or superconducting system, which in turn contains information on the
electronic structure of the pure system. The subset of the scattering wave
vectors \textbf{q} can be enhanced or not according to the type of disorder
and the phase structure of the superconducting gap, as noted by several groups
\cite{PhysRevB.73.104511,PhysRevB.68.180506} and experimentally verified in
the cuprate superconductor, Ca$_{2-x}$Na$_{x}$CuO$_{2}$Cl$_{2}$%
\cite{Hanaguri923,PhysRevB.80.144514}, where it was shown that QPI intensity
at certain \textbf{q} was selectively enhanced or suppressed by the external
magnetic field, consistent with $d_{x^{2}-y^{2}}$-wave pairing. Later the same
experiment was also performed for the FeSe$_{0.4}$Te$_{0.6}$
\cite{Hanaguri474}, although the interpretation of the results was less
obvious due to the absence of observable vortices in this particular
experiment.\cite{Hanaguri474,Sykora2011,QPI2015,MazinSingh2010} This
methodology is very efficient if an isolated set of features can be seprated
and monitored, as in the nodal $d$-wave symmetry. However, in multiband
systems with complicated momentum dependence of the order parameter, it boils
down to comparing mutiple theoretically-denerated complex patterns with the,
usually noisy, experimental picture.

Recently, we proposed a novel method of analyzing the QPI data from
non-magnetic impurities, which does not rely on the momentum dependence of
individual features, but rather upon integrated indicators, averaged, formally
over the entire momentum space. This method utilizes valuable phase
information, discarded in usual treatments  (see, however, Refs. \onlinecite{Bonn:2017i,Bonn:2017ii}. It was shown
that some integrated indicators, specifically the antisymmetrized combination
of the conductance, behave qualitatively differently for scattering events
accomponied or not by a change in the order parameter sign. This method does
not require manipulating scattering by magnetic field\cite{QPI2015}, whose
role as creator of artificial vortex \textquotedblleft
disorder\textquotedblright\ is unclear. The derivation in Ref. \onlinecite{QPI2015}
does not call for any assumption about the structure of the order parameter in
the momentum space, as long as it either has or does not have regions with the
opposite gap signs. While this derivation was completely general, it was also
confirmed by extensive numerical simulations with finite
disorder\cite{Martiny17}. Using this method, the authors of Ref.
\onlinecite{Sprau17} were able to identify a sign changing order parameter in
FeSe, and similar conclusions were recently drawn for Li hydroxide
intercalated FeSe as well\cite{Du2017}.

In this connection, a question arises whether a more detailed analysis of the
same characteristic, namely the antisymmetrized conductance, can be used to
extract more information than just an overall sign change. LiFeAs, for which
inetresting possibilities, such as sign-flipping between the hole Fermi
surfaces, have been suggested\cite{Ahn2014}, is  therefore a natural candidate. With this
goal in mind, we have extended our previous analysis to a realistic multiorbital
model of LiFeAs. The electronic structure of LiFeAs was previously measured by
ARPES and fitted to a 10 orbital tight-binding Hamiltonian\cite{Wang2013}.
This provides rather comprehensive information on the orbital and band
structure of this material in the low-energy regime. Furthermore, the
superconducting gap values for each of the Fermi surface sheets were also
measured both by ARPES\cite{sym4010251,PhysRevLett.108.037002} and previous
STM
measurements\cite{PhysRevLett.109.087002,PhysRevLett.108.127001,Allan563,PhysRevB.89.104522}%
. This provides an ideal testing ground for our earlier proposal given the
high-quality STM data available for
LiFeAs\cite{PhysRevB.86.174503,PhysRevB.94.134515,1703.07002}. Note that for
this material, the problem of the sign change of the gap between electron and
hole pockets was studied earlier by Chi et al.\cite{PhysRevLett.109.087002}
using ad hoc expressions for the LDOS based on BCS coherence factors, which
were shown in Ref. \onlinecite{QPI2015} to be incorrect; nevertheless similar
conclusions to [\onlinecite{QPI2015}] were reached. In addition, an attempt
was made to provide a quantiitative calculation of the STM spectra for various
impurities such as Mn, Ni or Co and different gap
symmetries\cite{PhysRevB.94.224518}. In this manuscript, we present a
qualtitative analysis for realistic band structure, using the correct
observable sensistive to gap sign change, but independent of the details of
impurity wave functions.

As expected, we find that despite the complications introduced by the multiple
Fermi surfaces, the distinction between the $s_{++}$%
and $s_{+-}$ can de readily made based on the integrated antisymmetrized
conductance. We further investigate how various symmetries, discussed in the
literature, can be identified via characteristic \textbf{q-}behavior of the
antisymmetrized conductance and could be used for a finer elucidation of the
gap sign structure in iron based materials up to detailed identification of
the gap sign corresponding to different pockets. Specifically, it is tempting
to separates electron-hole and hole-hole scattering features, and investigate
the latter for potential signature of the sign change. As we show below, while
such a separation is possible, proximity of the two gaps in the two main hole
bands precludes a firm conclusion about possible sign change between them.

\section{ Normal state band structure and QPI}

\begin{figure}[ptbh]
\renewcommand{\baselinestretch}{.8}
\includegraphics[angle=0,width=1\linewidth]{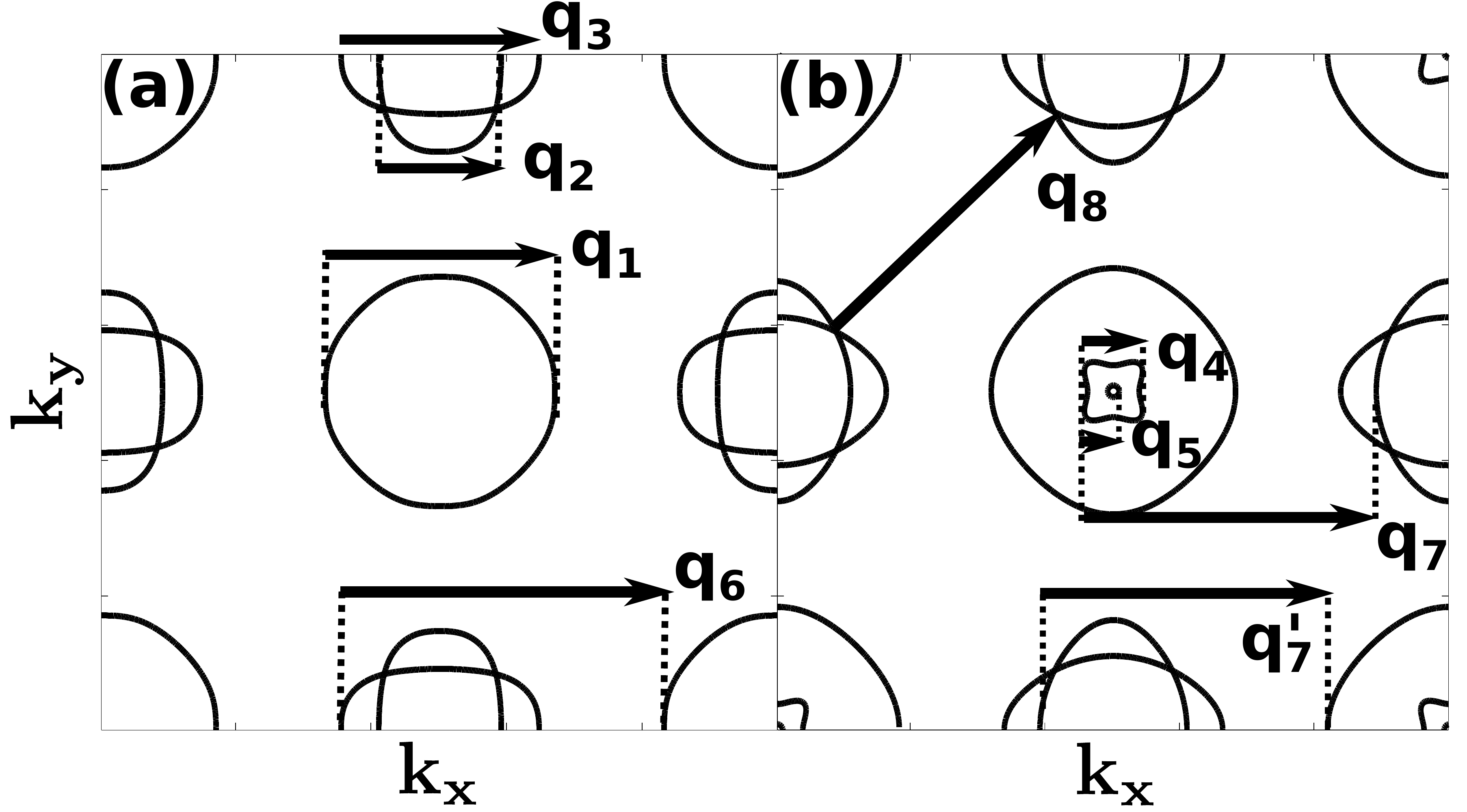}\caption{Cuts of
constant energy in LiFeAs for (a) $k_{z}=0$, and (b) $k_{z}=\pi$, plotted in
the 1-Fe Brillouin zone. Some potential scattering vectors $\mathbf{q}_{i}$are
indicated by the arrows. Chemical potential chosen to (a) $\mu=3  \text{ meV}$
and (b) $\mu=1 \text{ meV}$ corresponding to $n=6$ electrons in the system. }%
\label{fig:LiFermi}%
\end{figure}

We employ the ten-orbital tight-binding model Hamiltonian fitted to the ARPES
measured band structure of LiFeAs samples by Wang et al.\cite{Wang2013}, where
each 5 orbitals correspond to one of the two Fe-atoms within the unit cell.
The doping level of the system corresponds to $n=6$ $e^{-}$/Fe and the
resulting Fermi surface is shown in Fig. \ref{fig:LiFermi}(a) and (b) for two
values of $k_{z}=0$ and $k_{z}=\pi$. Due to the smallness of the Fermi
energies for the small hole pockets of the $xz,yz$-character they appear to
have stronger $k_{z}$ dependence and close around the Z point of the Brillouin
zone.\newline

The bare non-interacting Green's function is given by
\begin{equation}
\hat{G}^{0}(\mathbf{k},\omega)=\left(  i\omega_{n}-\hat{H}_{0}\right)
^{-1}.\label{bareG}%
\end{equation}
where $\omega_{n}$ refers to the Matsubara frequency. All quantities may be
considered to be matrices in band or orbital space, but the unperturbed
Green's function is diagonal in band space. The local density of state (LDOS)
is then found as $\rho(\omega)=-\frac{1}{\pi}\operatorname{Tr}%
\operatorname{Im}\sum_{\mathbf{k}}\hat{G}^{0}(\mathbf{k},\omega)$ where the
$\operatorname{Tr}$ runs over the orbital or band index. In the presence of
impurities the equation determining the full Green's function reads
\begin{equation}
\hat{G}(\mathbf{k},\mathbf{k}^{\prime},\omega)=\hat{G}^{0}(\mathbf{k}%
,\omega)+\hat{G}^{0}(\mathbf{k},\omega)\hat{t}(\mathbf{k},\mathbf{k}^{\prime
},\omega)\hat{G}^{0}(\mathbf{k}^{\prime},\omega),\label{FullG}%
\end{equation}
where the last term describes multiple scattering on impurities. In the
following we consider multiple scattering from a single pointlike spherically symmetric nonmagnetic
impurity \cite{Capriotti2003,Zhu2004}, i.e. the impurity potential $\hat{U}$
is taken to be well-localized and independent of spin. The $t$-matrix is then
no longer momentum dependent, and is related to the impurity potential by
${\hat{t}}={\hat{U}}+{\hat{U}}{\hat{G}^{0}}{\hat{t}}$.

In such a case the solution for the $t$-matrix may be written,
\begin{equation}
{\hat{t}(\omega)}=[1-{\hat{U}}\sum_{\mathbf{k}}{\hat{G}(\mathbf{k},\omega
)]}^{-1}{\ {\hat{U}}}.
\end{equation}
Thus the position dependent correction to the LDOS reads may be written
$\rho(\mathbf{q},\omega)=\rho(\omega)+\delta\rho(\mathbf{q},\omega)$ with
\begin{equation}
\delta\rho(\mathbf{q},\omega)=-\frac{1}{\pi}\mathrm{Tr}\,\operatorname{Im}%
\sum_{\mathbf{k}}\hat{G}^{0}(\mathbf{k},\omega)\hat{t}(\omega)\hat{G}%
^{0}(\mathbf{k}+\mathbf{q},\omega)\label{drhoFourier}%
\end{equation}
where in the last equation we employ the analytic continuation to the advanced
(retarded) Green's functions.

\begin{figure}[ptb]
\renewcommand{\baselinestretch}{1}
\includegraphics[angle=0,width=1\linewidth]{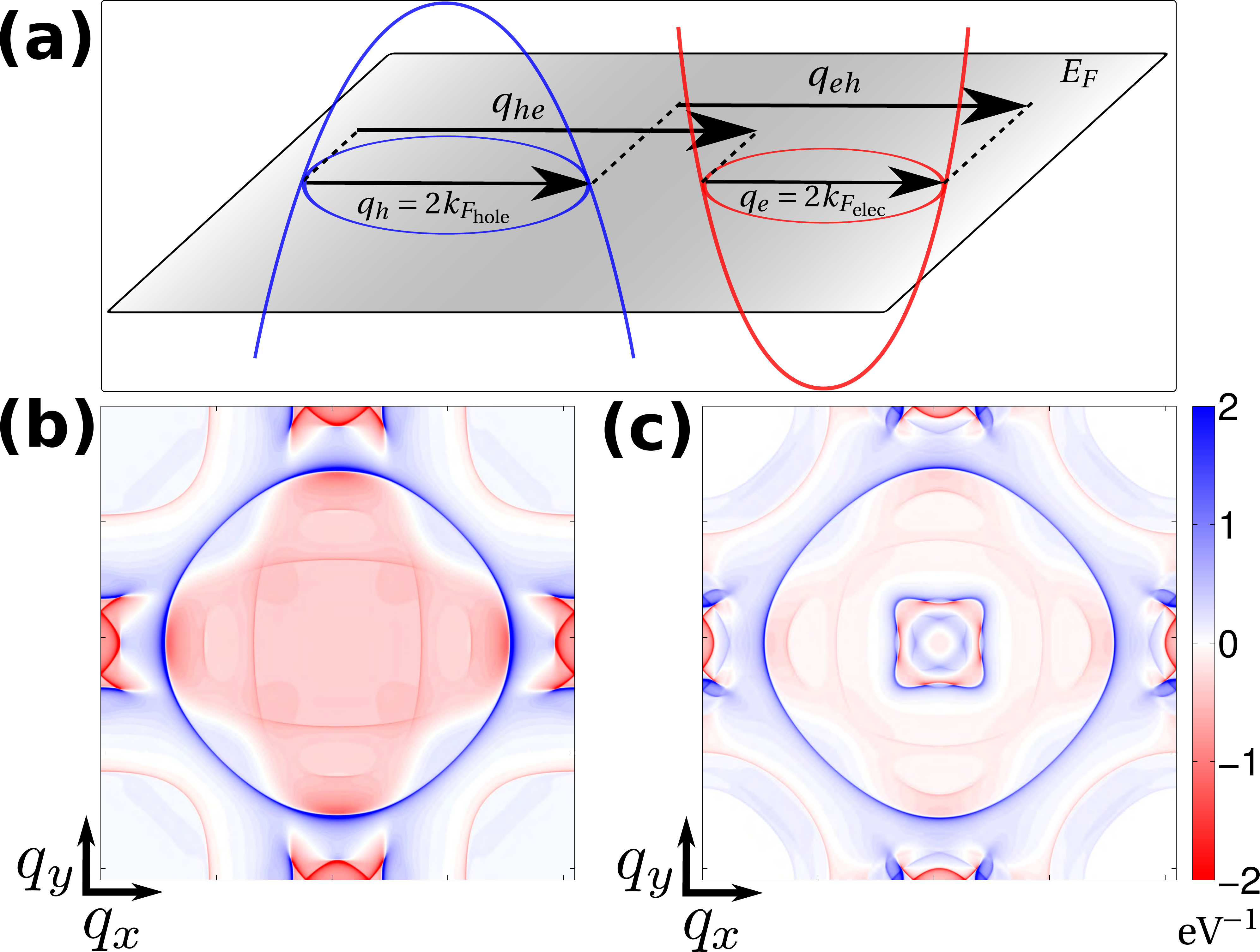}\caption{(a) Schematic representation of possible scattering events in a one electron and one hole pocket system. (b),(c)  Quasiparticle
interference (momentum dependent correction to the LDOS) $\delta
\rho(\mathbf{q},\omega=0)$ maps shown in the full Brillouin zone for Born
limit scattering in the orbital space $U=1 \text{ meV}$. Cuts (b) for $k_{z}=0$
and (c) $k_{z}=\pi$. The sign of the correction to the density of states for
the intraband scattering reflects whether the band is hole-like (positive) or
electron-like (negative) }%
\label{fig:Fullqpi}%
\end{figure}

We have not yet specified the form of the impurity potential in band or
orbital space. Intuition suggests and density functional theory calculations
confirm to a very good approximation that $\hat U$ is diagonal in orbital
space\cite{Nakamura2011}. In the following, until specified otherwise we
assume the impurity potential to be diagonal in the orbital space with equal
amplitudes for each orbital. The resulting normal state quasiparticle
interference (momentum-dependent correction to the local density of states)
map is shown at zero frequency in Fig. \ref{fig:Fullqpi} for the two
characteristic $k_{z}$ cuts and employing the Born approximation with a weak
scattering potential $U=1$ meV. Comparing the patterns to the ones found in
experiment in Ref. \onlinecite{PhysRevB.89.104522}, it appears that figure
\ref{fig:Fullqpi}(b) is in closest agreement, with well defined intraband
($q\approx0$) and interband ($q\approx(\pi,0))$ peaks.
%In both the clear structure of the intra pocket scattering is found while the inter pocket scattering is dominant at the area around $X$ and $Y$ point.
\footnote{Note that Ref. \onlinecite{PhysRevB.89.104522} has to be rotated by
45%
%TCIMACRO{\U{b0} }%
%BeginExpansion
${{}^\circ}$
%EndExpansion
and reduced to the first Brillouin zone.} 
The structure of the Fermi surface and the character of the
carriers can be easily deduced from the patterns. Once the sign of the
impurity potential is fixed (we assume it to be positive for clarity), the
scattering within hole or within electron pockets are clearly separable. While
the scattering within hole pockets at $q=2k_{F}$ is positive, it is negative
for the electron ones. This is determined by the initial sign of electronic
dispersion with respect to the chemical potential. The interband scattering between
electron and hole pockets contains both positive and negative contributions to $\delta\rho$. In particular, there are two processes of opposite signs,  see Fig. \ref{fig:Fullqpi} (a). The first one is hole-like (positive) and another one is electron-like (negative). Observe that  if the hole pocket is larger than the electron one, the hole-like scattering can appear in the  second BZ, while the electron-like is in the first BZ, and vice versa if the pocket sizes are reversed. Thus the interband peaks in Fig.\ref{fig:Fullqpi}{\red  (b),(c) } can be either positive or negative depending on the relative size and ellipticity of the pockets.

\begin{figure}[ptb]
\renewcommand{\baselinestretch}{.8}
\includegraphics[angle=0,width=1\linewidth]{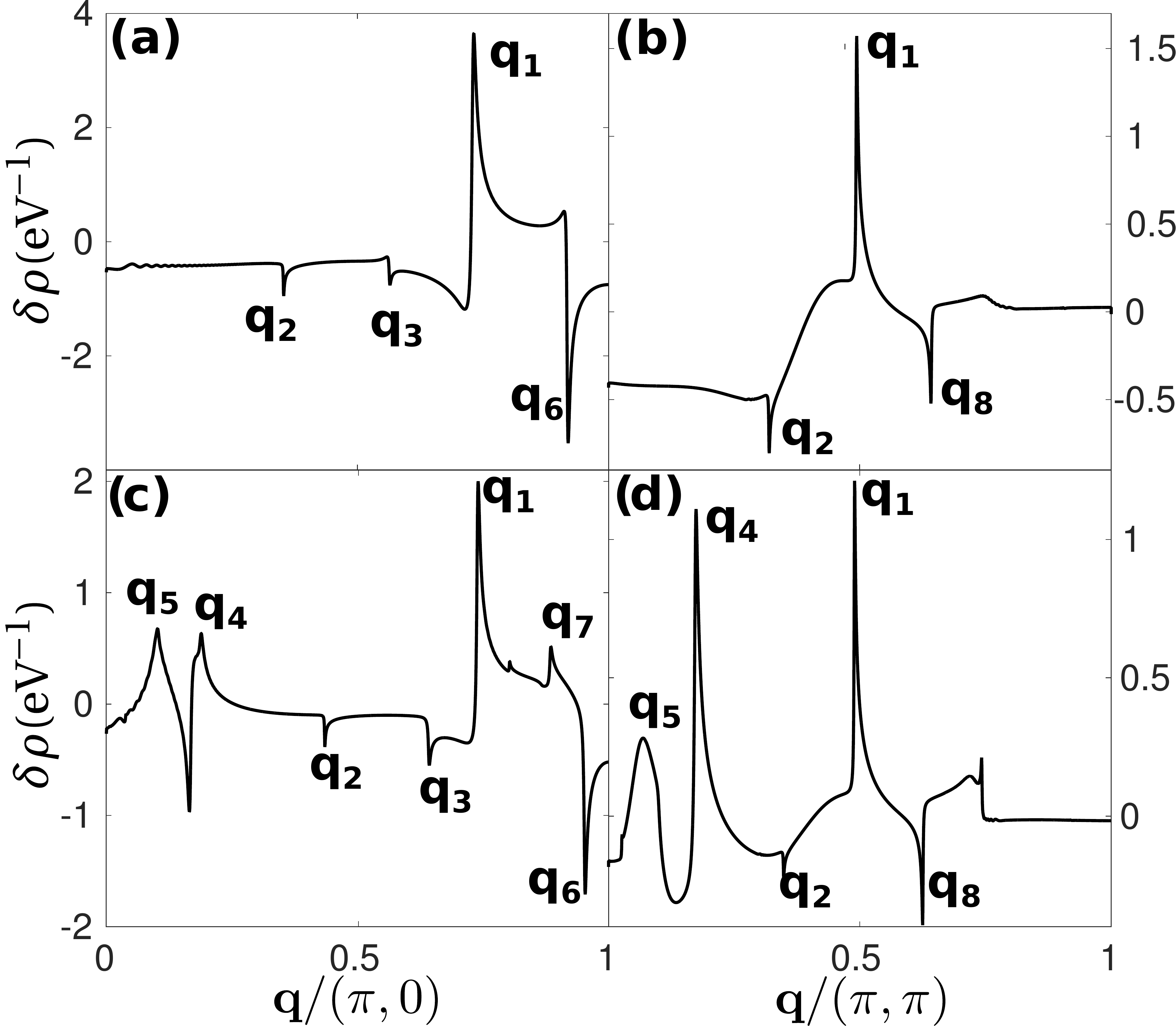}\caption{Cuts
taken from figure \ref{fig:Fullqpi}. $\delta\rho(\mathbf{q},\omega=0)$
calculated along the symmetry routes $\Gamma\to X$ (left panel) and $\Gamma\to
M$ (right panel) in the first Brillouin zone for $k_{z}=0$ (a),(b) and
$k_{z}=\pi$ (c),(d). The corresponding scattering wavevectors at the Fermi
surface cuts are shown in figure \ref{fig:LiFermi}.}%
\label{fig:Likz0}%
\end{figure}

We first examine the QPI peaks that occur due to the complicated Fermi surface
in the normal state. The strongest contribution to the LDOS is given by
scattering between like orbitals within the large hole pocket (of
$xy$-character), as well as within the small hole pocket ($xz,yz$-character).
The interband scattering between electron and hole pockets, again of $xz$,$yz
$-character (Fig. \ref{fig:LiFermi}), is also significant. To visualize the
corresponding structure of the scattering peaks we present in Fig.
\ref{fig:Likz0} the $\delta\rho(\mathbf{q},\omega)$ along the $\Gamma\to X$
and $\Gamma\to M$ high symmetry paths of the first BZ for two different $k_{z}
$ cuts. Observe that all the scattering events can be clearly identified
with processes corresponding to either intraorbital intraband and interband
scattering. In particular, as shown in Fig. \ref{fig:LiFermi}, scattering
vectors $\mathbf{q}_{1}$ till $\mathbf{q}_{4}$ refer to the intra pocket
scattering and the amplitude of intra hole and intra electron pocket
scattering peaks differs in sign. While the correction to the LDOS from the
scattering on the electronic band has a negative sign, the intrapocket
scattering for the hole-like bands is positive (for assumed $V_{imp}>0$; all
signs reverse for the opposite case in the weak scattering limit). This
difference is related to the overall sign of the electron and hole dispersion.
This information can be used to identify interpocket scattering vectors
between the pockets of the same character. For example, $\mathbf{q}_{5} $
refers to the scattering between the small hole pockets, while $\mathbf{q}%
_{8}$ is the scattering between two electron pockets located at $X$ and $Y$
point of the BZ. For interband scattering we identify the scatterings between an
electron pocket and the large hole pocket which are labeled as $\mathbf{q}%
_{6}$ and $\mathbf{q}_{7}$ by examining the
corresponding energy cuts in the first BZ.

Once the scattering events in the Born limit are understood, we should mention
that by increasing the impurity scattering strength we find within T-matrix
that there are also interorbital scattering components introduced for example
between larger and smaller hole pockets, which are however still weaker than
the intraorbital ones. In the following, however, we proceed directly to the
superconducting state.

\section{Superconducting state}

Superconducting gap magnitudes and their angular variations on  the electron
and hole pockets are known for LiFeAs from  ARPES
\cite{PhysRevLett.108.037002,sym4010251,Wang2013} and STM\cite{Allan563}
measurements yet their relative phases are unknown. In particular, the
magnitude of the superconducting gap for the small hole pockets is $\Delta
_{h}=5 \text{ meV}$, for the large hole pocket $\Delta_{H}=2.5 \text{ meV}$ and
for the electron pockets $\Delta_{e} \approx3\text{ meV}$. These measurements
do not fix the gap structure completely, however, because the relative phases
are unknown. There have been a number of theoretical proposals regarding the
phase structure of the superconducting gap in LiFeAs. In the original analysis
using spin-fluctuation mediated Cooper-pairing performed by Wang et al.
\onlinecite{Wang2013} using the tight-binding band structure discussed above,
the usual $s_{+-}$-wave symmetry of the superconducting gap with an overall
change of sign between electron and hole pockets was found.  On the other hand, the calculated gaps on the
small Z-centered pockets were  too small compared to experiment. 
Several subsequent works explored ways to cure this
discrepancy\cite{Ahn2014,PhysRevB.94.201109,Kontani2014}. In Ref.
\cite{Ahn2014}, it was pointed out that once the spin fluctuations are
relatively weak, which seems to be the case in LiFeAs\cite{PhysRevB.90.144503}%
, there are several $s$-wave channels which are nearly
degenerate\cite{Ahn2014}. Most importantly, these $s$-wave states may involve
sign changing gap on the two hole pockets\cite{Ahn2014} as well as
orbital-antiphase $s-$wave gap\cite{PhysRevB.94.201109}, which is stabilized
when the pairing interaction is closer to being diagonal in orbital space than
band space\cite{Yin2014}.

Note that all of these works claim to be roughly consistent with existing
ARPES experiments, but find different sign structures of the gaps on the
various Fermi surface sheets. In the following, we compute the QPI signatures
of each of the $s-$wave states, which should help to identify the particular
symmetry of the order parameter in LiFeAs. In real STM measurements, electrons
from a range of $k_{z}$ states are involved in the tunneling process, so no meaningful
choice of a single $k_{z}$-cut is possible. In 2D systems like cuprates, the
influence of the third momentum dimension is negligible, but in LiFeAs this is
not the case (Fig. \ref{fig:LiFermi}). While electron and large hole pockets correspond to bands that are barely dispersing in the $k_{z}$-direction, the same is not true for he small hole pockets.
Nevertheless, empirically the QPI data on LiFeAs seem to
contain large scattering vectors that connect the small hole pockets with
the electron pockets, suggesting that the $k_{z}=\pi$ states play an essential
role.  With this in mind, 
we continue our calculation with $k_{z}=\pi$ where all
pockets are visible on the Fermi surface.\cite{1710.02276}

\begin{figure}[ptbh]
\includegraphics[width=1\linewidth]{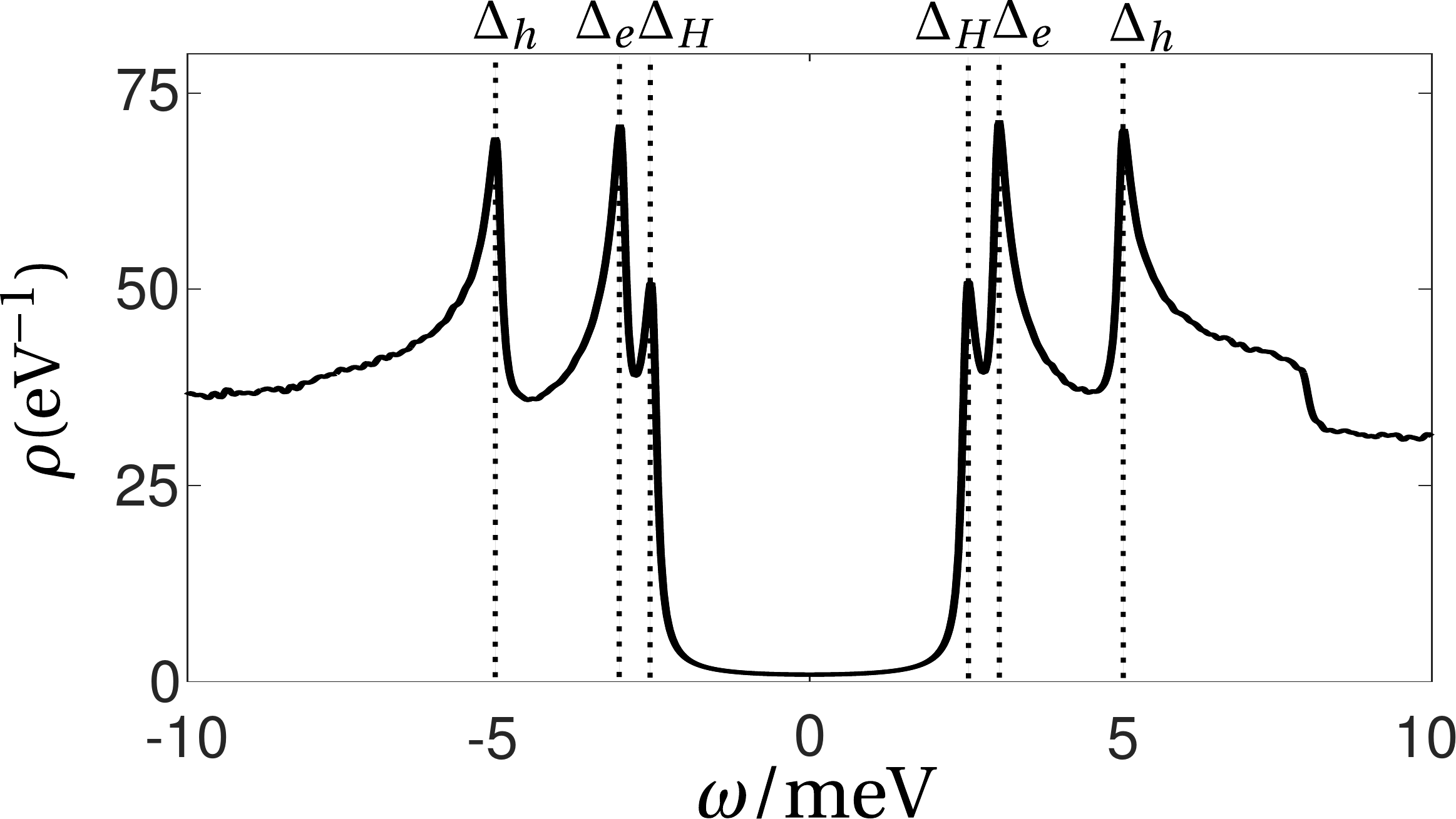}\caption{Density of states in the
superconducting phase $\rho(\omega)$. Integrated over the full Brillouin zone
and in presence of all three hole pockets at $k_{z}=\pi$. Vertical dashed
lines indicating gap values $\Delta_{h}=5 \text{ meV}$, $\Delta_{H}%
=2.5 \text{ meV}$ and $\Delta_{e}=3 \text{ meV}$.}%
\label{fig:DOS}%
\end{figure}

In the superconducting state, Eq.
\ref{drhoFourier} is generalized to
\begin{equation}
\rho(\omega)=-\frac{1}{2\pi}\operatorname{Tr}\operatorname{Im}\sum
_{\mathbf{k}}(\tau_{0}+\tau_{3})\hat{G}^{0}(\mathbf{k},\omega)\quad,
\end{equation}
where $\tau_0$ and $\tau_1$ are  Pauli matrices in Nambu (particle-hole) space.
We show the total density of states in Fig. \ref{fig:DOS} for the gap structure
($\Delta_{h},\Delta_{H},\Delta_{e}$) as determined by ARPES. A typical $U$-shaped spectrum is found with all three gaps visible. All the gaps are well pronounced in the spectrum.
At energies above the coherence peaks,  normal state properties are recovered, showing a weak particle
hole asymmetry. Hence, the antisymmetric part of the correction to the LDOS can be
found as
\begin{equation}
\delta\rho^{-}(\mathbf{q},\omega)=\mathrm{Tr}\,\operatorname{Im}%
\sum_{\mathbf{k}}\tau_{3}\hat{G}^{0}(\mathbf{k},\omega)\hat{t}(\omega)\hat
{G}^{0}(\mathbf{k}+\mathbf{q},\omega),\label{drhoFouriersc}%
\end{equation}
where the $t$-matrix in the Nambu space is given by
\begin{equation}
{\hat{t}(\omega)}=[1-\tau_{3}{\hat{U}}\sum_{\mathbf{k}}{\hat{G}(\mathbf{k}%
,\omega)]}^{-1}{\tau_{3}{\hat{U}}}\label{Tmat}%
\end{equation}
and $\tau_{3}$ is the corresponding component of the Pauli matrix. In the
following we will consider various representation of the phases of the
superconducting gap on the Fermi surface pockets. However, we will also study the
role of the impurity potential in changing the characteristic features of the
$\delta\rho^{-}(\mathbf{q},\omega)$.

\subsection{$s_{+-}$-wave gap: impurity potential in orbital and band bases}

\begin{figure}[ptbh]
\includegraphics[width=1\linewidth]{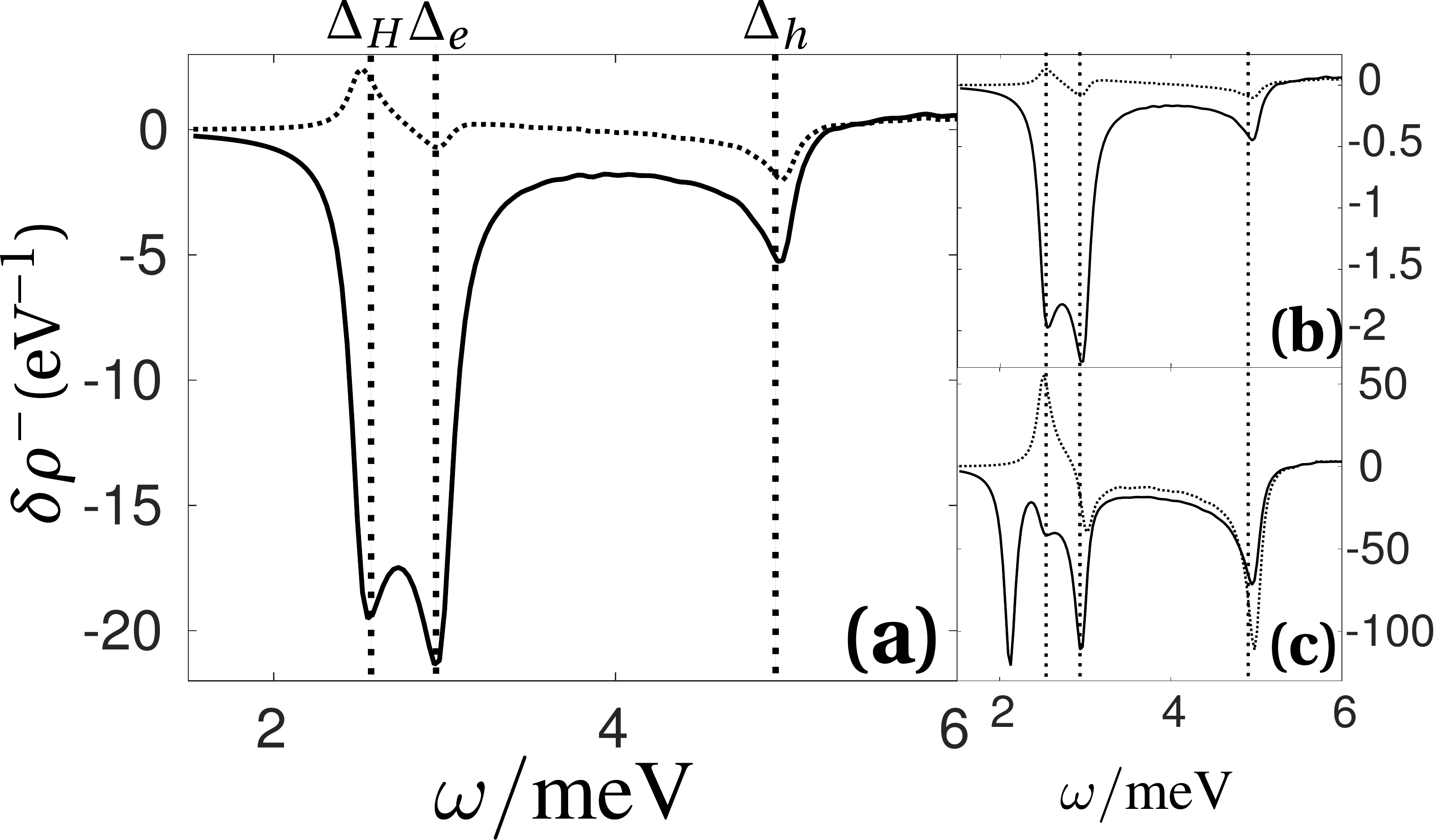}\caption{Antisymmetrized
$\mathbf{q}$ integrated correction to the LDOS $\delta\rho^{-} (\omega)$ for
non-magnetic impurity scattering diagonal in orbital space. (a) intermediate
limit full $t$-matrix calculation $U=10 \text{ meV}$, (b) Born approximation
$U=1 \text{ meV}$ and (c) unitary limit full $t$-matrix calculation
$U=0.2\text{eV}$ integrated over the full Brillouin zone and in presence of
all three hole pockets at $k_{z}=\pi$. Vertical dashed lines indicating gap
values. Result is shown with solid lines for conventional $s_{+-}$, dashes
lines for $s_{++}$.}%
\label{fig:Fullqint}%
\end{figure}

Before we start our analysis by studying the different possible gap
structures, we need to point out the importance of the actual model used for
the impurity. First, we consider an impurity which is diagonal in orbital
space, $V_{imp}^{\ell_{1}\ell_{2}} = U\delta_{\ell_{1}\ell_{2}}$, and consider
the effect of the potential strength $U$ on our ability to identify sign-changing gaps. For simplicity, we first consider the usual $s^{\pm}$-wave
superconducting gap {\it i.e.}, the overall phase of the gap changes from 0 to $\pi$
between electron and hole pockets. In Fig.\ref{fig:Fullqint} we show the bias
dependence of the full momentum space integrated correction to the LDOS in the
Born, intermediate and strong scattering limit in the presence of all three
hole pockets obtained by
\begin{equation}
\delta\rho^{-}(\omega) =\sum_{\mathbf{q}\in\Omega} \delta\rho^{-}%
(\mathbf{q},\omega),\label{drhoint}%
\end{equation}
where the integration is over the entire momentum space, or, in practical application, over a well defined area $\Omega$ around the scattering peaks
\cite{QPI2015}.

Comparing the Born limit with $U=1 \text{ meV}$ shown in Fig.\ref{fig:Fullqint}%
(b) to the result of the intermediate strength scattering $U=10 \text{ meV}$
shown in figure \ref{fig:Fullqint}(a), we find them in close agreement with
each other. The clear  ``even'' behavior of $\delta\rho^{-}(\omega)$ for
the sign changing $s$-wave state and the ``odd'' one for the $s_{++}$ state
are expected.\cite{QPI2015} Here we use the
imprecise terms "even" and "odd" behavior of the antisymmetrized LDOS to refer,
respectively, to single-sign, resonant behavior of $\delta\rho^{-}$ as a function of bias
between the gap energies for the $s_{+-}$ state and the weaker response with
a zero-crossing for the $s_{++}$ state.\cite{QPI2015}

The characteristic signal changes as soon as we introduce stronger scattering
amplitudes $U=0.2\text{eV}$ in Fig. \ref{fig:Fullqint}(c). An in-gap bound
state develops for $s_{+-}$ symmetry, moving from the gap edge toward smaller
energies with increasing scattering strength. The occurrence of in-gap bound
states is of course itself a strong hint toward a sign change of the order
parameter; nevertheless the distinction between even and odd $\delta\rho^{-}$
signals in the bias region between the two gap energies remains unambiguous,
so the antisymmetrized LDOS analysis may also be used as a test of gap sign
change after subtracting any bound states \cite{QPI2015,Du2017}.

While for a general statement of the sign change this is sufficient, the
distribution of signs over the different pockets is not determined by this
procedure. 
LiFeAs, with its well studied Fermi surface, is a promising material for
addressing possible scattering events separately. In order
to use the known fermiology to
restrict our integration range to a well defined area of \textbf{q}-space and
study the response in an experimentally relevant case, we consider scattering
in band space, where we can select scattering events simply by including or
excluding specific pockets in the scattering.

\begin{figure}[ptbh]
\includegraphics[width=1\linewidth]{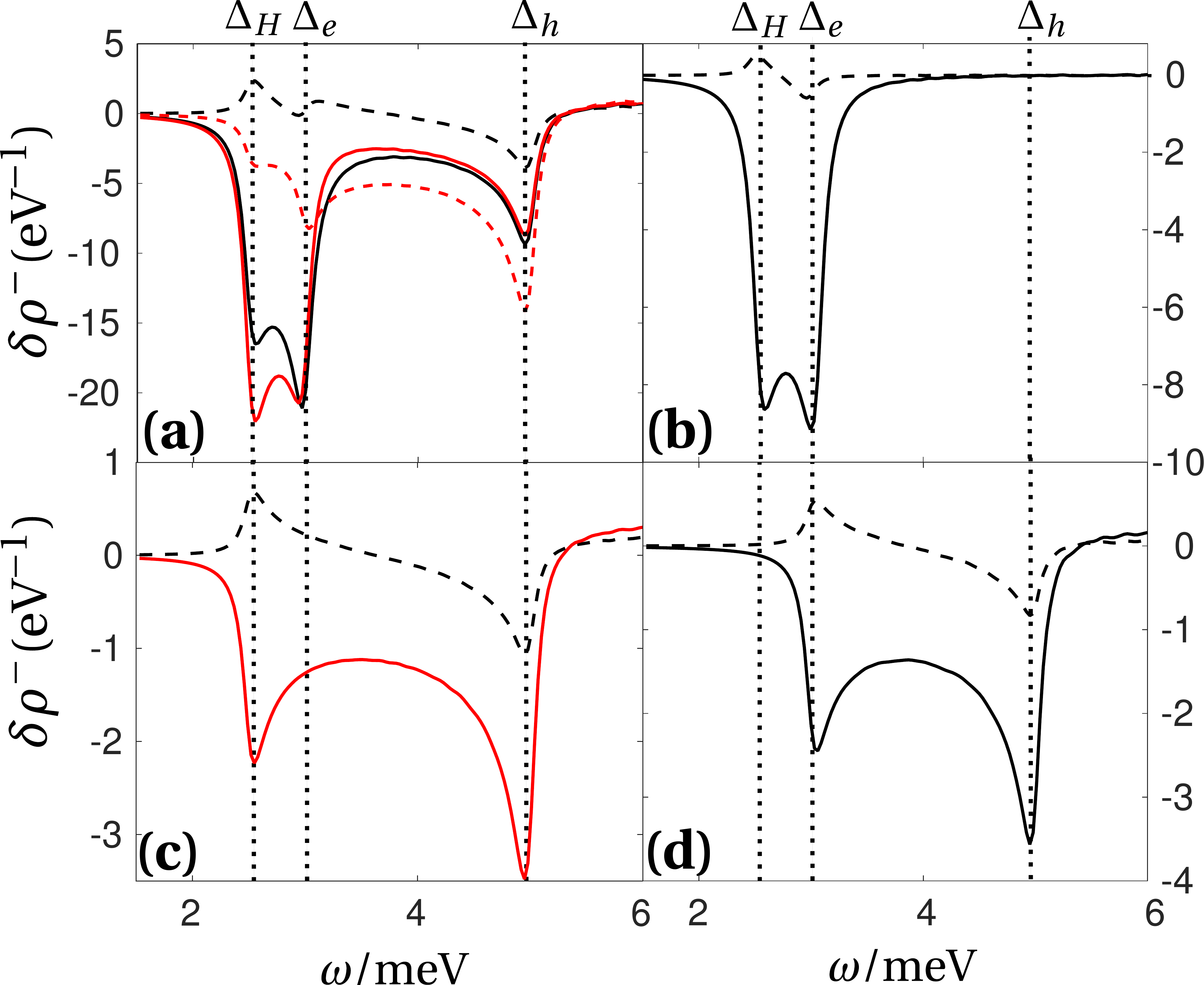}\label{fig:Bandsp}
\caption{Antisymmetrized $\mathbf{q}$ integrated correction to the LDOS
$\delta\rho^{-}(\omega)$ for non-magnetic impurity scattering potential
diagonal in band space and in Born limit $U_{\text{inter}}=U_{\text{intra}%
}=1 \text{ meV}$. (a) sum of all inter and intraband scatterings, (b) integrated
$q$ area containing large hole and electron pocket scattering, (c) integrated
$q$ area containing large hole and small hole pocket scattering
and (d) small hole and
electron pockets scattering. Dashed vertical lines indicate the magnitudes of
the corresponding gaps. Solid black curves refer to the conventional $s_{+-}%
$-state , with all hole pocket gap signs the same, and opposite to that on the
electron pockets (Fig. \ref{fig:antiphase}(B) The dashed black curves
correspond to $s_{++}$, the solid red line refer to the state containing an
sign change of the gap on the large hole pocket Fig. \ref{fig:antiphase}(C),
while the dashed red one corresponds to the sign change on small hole pockets
(Fig. \ref{fig:antiphase}(D).}\end{figure}

In particular, in Fig.
\ref{fig:Bandsp}(a) the $q$ integrated scattering is shown in the Born limit.

%We need to clarify that the overall increase of the magnitude of our response as compared to the diagonal impurity potential defined in the orbital space  is expected as we now include all matrix elements of $U$ and not only the diagonal ones.

Choosing $U_{\mathrm{inter}}=U_{\mathrm{intra}}=1 \text{  meV}$, we find a
similar behavior for the $s_{++}$ and $s_{+-}$ case to those shown previously
in Fig.\ref{fig:Fullqint} for the impurity potential defined in orbital space.
This indicates that the behavior is universal and does not depend on the way
the impurity potential is introduced.

\begin{figure}[ptb]
\centering
\includegraphics[width=0.7\linewidth]{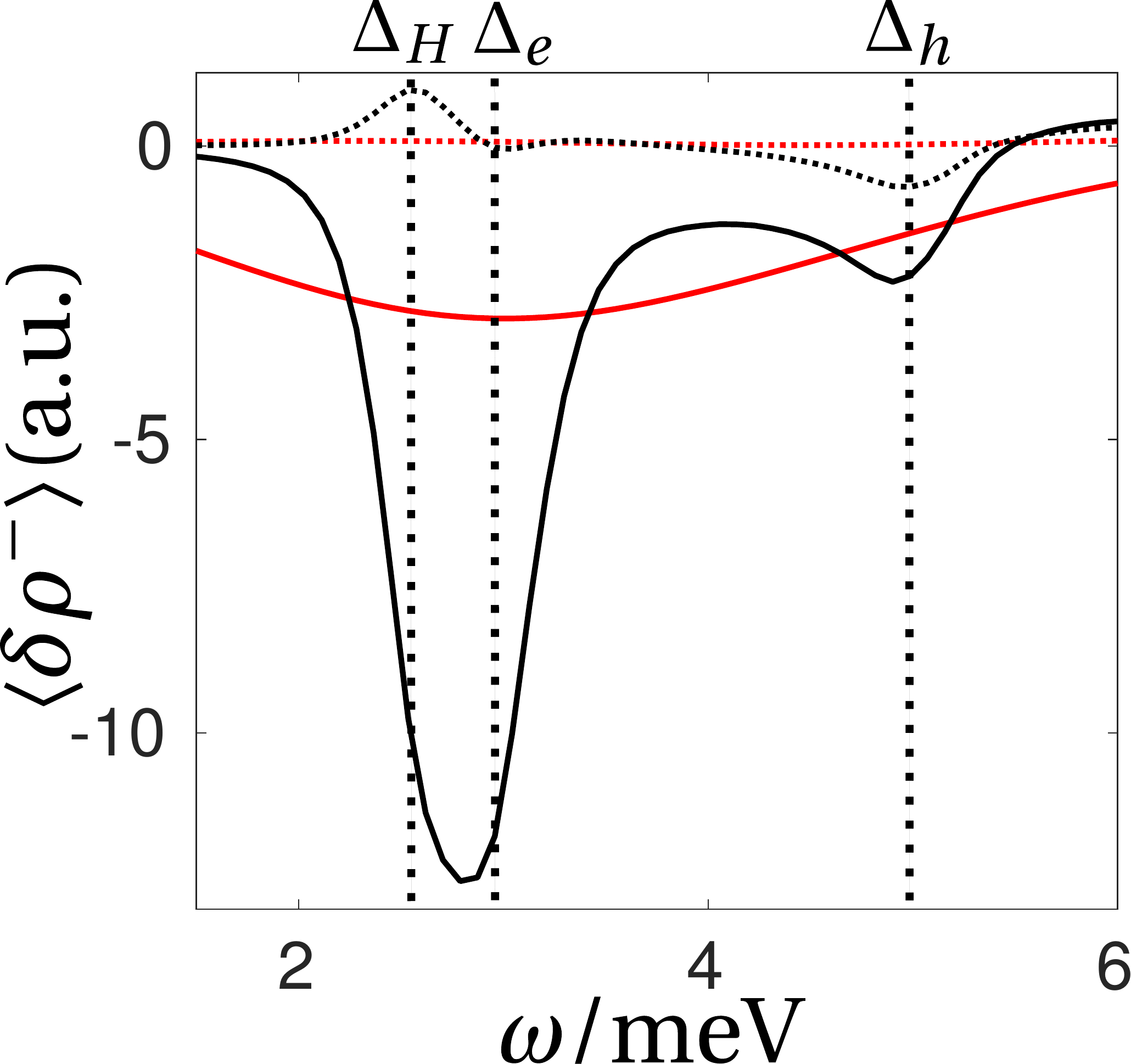}\caption{ Thermally averaged
antisymmetrized full-$\mathbf{q}$ integrated LDOS change $\langle\delta
\rho^{-}(\omega)\rangle$ for nonmagnetic scattering in the intermediate limit
$U=10$ meV (the $\delta\rho^{-}(\omega)$ is shown in Fig.5(a)) and $k_{z}=\pi$.
The black an red curves refer to $T=0.2T_{c}$ and $T=0.6T_{c}$, respectie.
Solid curves refer to the $s_{+-}$ state, while the dashed ones for
conventional $s_{++}$ state.}%
\label{fig:?l?l}%
\end{figure}

In realistic STM experiments, the measurements are done at finite yet very low
temperatures. As we have shown previously in Ref. \onlinecite{QPI2015}, a
thermal broadening smears out the clear peak structure with increase of
temperature. Nevertheless, as we show in Fig. \ref{fig:?l?l}, the
characteristic features of the $s_{+-}$ and $s_{++}$-states remains still
visible up to $T=0.6T_{c}$. Here, the temperature dependence is introduced
using BCS-type behavior of the gap magnitudes on each Fermi surface pocket and
thermally averaging the antisymmetrized correction to the local density of
states
\begin{equation}
\langle\delta\rho^{-}(\Omega)\rangle\equiv\int_{-\infty}^{\infty}d\omega
\delta\rho^{-}(\omega)\left[  {\frac{-\partial f}{\partial\omega}}%
(\omega+\Omega)-{\frac{-\partial f}{\partial\omega}}(\omega-\Omega)\right]
\nonumber\label{eq:Tdep}%
\end{equation}

\subsection{Other possible $A_{1g}$-symmetry states with sign-change gap
between hole pockets}

\begin{figure}[ptbh]
\includegraphics[width=0.9\linewidth]{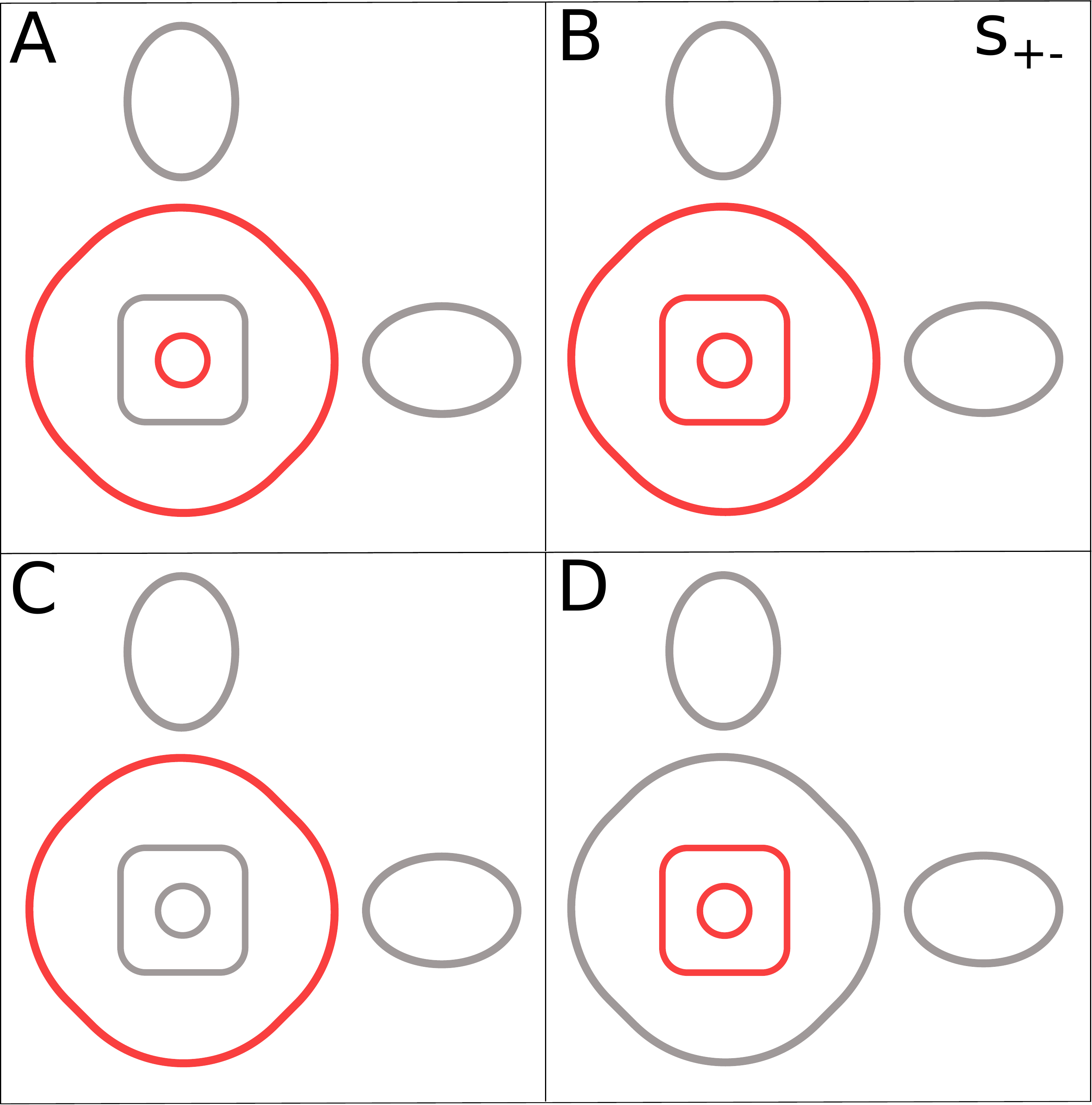}
\caption{Possible sign-changing states within $A_{1g}$-symmetry representation
found previously in Ref.\cite{Ahn2014}. The conventional $s_{+-}$-state is
referred as $B$-state. Red and grey indicate
different signs of the gap function on the Fermi surface.  Note we do not discuss the interesting further possibility that there might be complex phases among the gap functions on the different Fermi sheets, as predicted in some cases\cite{Maiti2013}.  This question has been addressed in Ref. \onlinecite{Boeker2017}, where it was found that such complex states did not yield clear qualitative signatures in $\delta \rho^{-}$.}
\label{fig:antiphase}%
 \end{figure}

Due to the relative weakness of the spin fluctuations at the wave vector
corresponding to the scattering between electron and hole pockets in
LiFeas\cite{Qureshi2012}, it has been proposed\cite{Ahn2014} that the
distribution of the phases of the superconducting gap magnitudes on the Fermi
surface sheets could be different than it is in the conventional $s_{+-}$-wave
scenario. Overall the gap would still possesses the global $A_{1g}$-symmetry,
yet the phase structure could be richer. Some such states are depicted in
Fig.\ref{fig:antiphase}. 

The
natural question arises whether QPI could be used to determine these
additional phase shifts. We first note that the $A$-state has an additional
sign change between two tiny hole pockets, located near the $\Gamma$-point.
For the QPI it would mean to search for the characteristic features in
$\delta\rho^{-} (\omega)$ for  small scattering wavevectors ($\mathbf{q}%
_{4}$ and $\mathbf{q}_{5}$), which is in general difficult for the QPI
experiments as it requires a Fourier transform of Friedel oscillations from an
isolated impurity recorded in a large field of view in the real space. In
addition these small pockets exist only near $k_{z} = \pi$ and have much
smaller densities of states, which represents an additional problem.
Therefore, we first address the question of whether  some less subtle possible
sign changes beyond the usual $s_{+-}$ picture can be still detected. We plot in Fig.
\ref{fig:Bandsp}(a) the results for the only two other candidate states for
LiFeAs, namely for the superconducting state where the large hole pocket of
mostly $xy$-chraracter has opposite sign to all other electron and hole
pockets ($C$-state) and the one where the tiny inner hole pockets have
opposite sign to the large hole and electron pockets ($D-$state). { We find  an even
response  for the total
integrated $\delta\rho^{-} (\omega)$, for each sign-changing gap structure,
when the impurity scattering potential is written in the band basis}. Thus, it appears that the magnitude of
the response at the different gap values provides direct information regarding the
pockets in which the sign change occurs. To specify this further we show in Fig.
\ref{fig:Bandsp} (c)-(d) the integrated responses around each corresponding
$q$ vector i. e. large hole to electron, small hole to electron and hole to
hole scattering, respectively. All three possible scattering events show
indeed the expected even or odd symmetry when the change of the sign of the
gap occurs.

To state confidently whether $C$ and $D$ states can be identified, we now
return to a more realistic situation, 
when the impurity potential is   nearly
diagonal in the orbital basis. This seems to agree better with ab-initio
calculations\cite{Nakamura2011,PhysRevB.94.224518}. The problem is further
complicated by the fact that beyond the weak scattering (Born) limit, the
$t$-matrix introduces additional interorbital scattering terms. Here we study
the effects of such terms by considering an intermediate strength
scatterer $U=10$ meV that does not create bound states in the gap.
\begin{figure}[ptb]
\renewcommand{\baselinestretch}{.8}
\includegraphics[angle=0,width=1\linewidth]{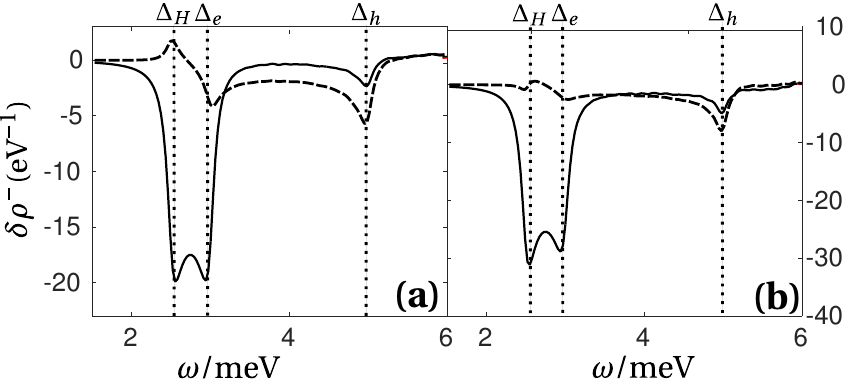}\caption{
Antisymmetrized $\mathbf{q}$ integrated correction to the LDOS $\delta\rho^{-}
(\omega)$ for the $C$ and $D$-states, depicted in Fig.7, using the
non-magnetic impurity diagonal in orbital space using $t$-matrix calculation
for $U=10 \text{ meV}$ in presence of all three hole pockets at $k_{z}=\pi$ .
\textbf{(a)} refers to full $\mathbf{q}$ integration of the QPI map as also
shown Fig.\ref{fig:Fullqint}(a). \textbf{(b)} shows $q$-selected scattering
around $\mathbf{q}_{7}$. Solid curves refer to the $C$-state (sign change on
the large hole pocket), while the dashed curves refer to the $D$-state.}%
\label{fig:wdepFullqandq7}%
\end{figure}In Fig.\ref{fig:wdepFullqandq7} (a) we show the antisymmetrized
response integrated over the full $q$ space for the $C$ and the $D$-states. We
remind the reader that for the $s_{++}$ and the conventional $s_{+-}$ states,
the results are clearly separable independent of the way the impurity
potential is written. Regarding the more complicated $C$- and the $D$-states,
we note that the relative sign change ($C$-state) or its absence ($D$-state)
between large hole and electron pockets, i.e. between $\Delta_{H}=2.5$ meV and
$\Delta_{e}=3$ meV is clearly visible. At the same time, the variation of $\delta\rho^-$ between
smaller gaps, $\Delta_{e}$ and the large gap, $\Delta_{h}=5$  meV, on the small
pockets is trickier due to a more complex phase structure of the overall gap.
While for the $D$-state the clear sign-changing behavior is indeed observed,
as expected, the behavior of $\delta\rho^{-}$ cannot be clearly assigned for
the $C$-state. In the latter state, $\Delta_{e}$ has opposite sign to
$\Delta_{H}$, but the same as on $\Delta_{h}$, so the behavior of $\delta
\rho^{-} (\omega)$ loses its characteristic behavior between $\Delta_{e}$ and
$\Delta_{H}$. The situation does not really improve if one selects some
specific $q$-scattering wavevector such as $q_{7}$, which is responsible for
the scattering between small hole pockets and electron pockets. The results in Fig. \ref{fig:wdepFullqandq7} (b) again show a clear signature of the
phase structure between $\Delta_{H}$ and $\Delta_{e}$, but cannot be unambiguously
assigned for frequency region between $\Delta_{e}$ and $\Delta_{h}$.

The result of this exercise shows that the determination of the
superconducting gap structure on  the multiple Fermi surface sheets may be a
challenging task if  gaps change sign more than once ({\it i.e.}, a
frustrated case), or if some gaps are close in magnitude.
For this situation, $\delta\rho^{-}(\omega)$ can be still
helpful to identify whether the gap is generally sign changing at least on
some of the pockets, however, the overall phase structure distribution over the
mutiple sheets cannot be unambigously determined.

\subsection{Orbital anti-phase superconductivity}

\begin{figure}[ptb]
\includegraphics[angle=0,width=0.7\linewidth]{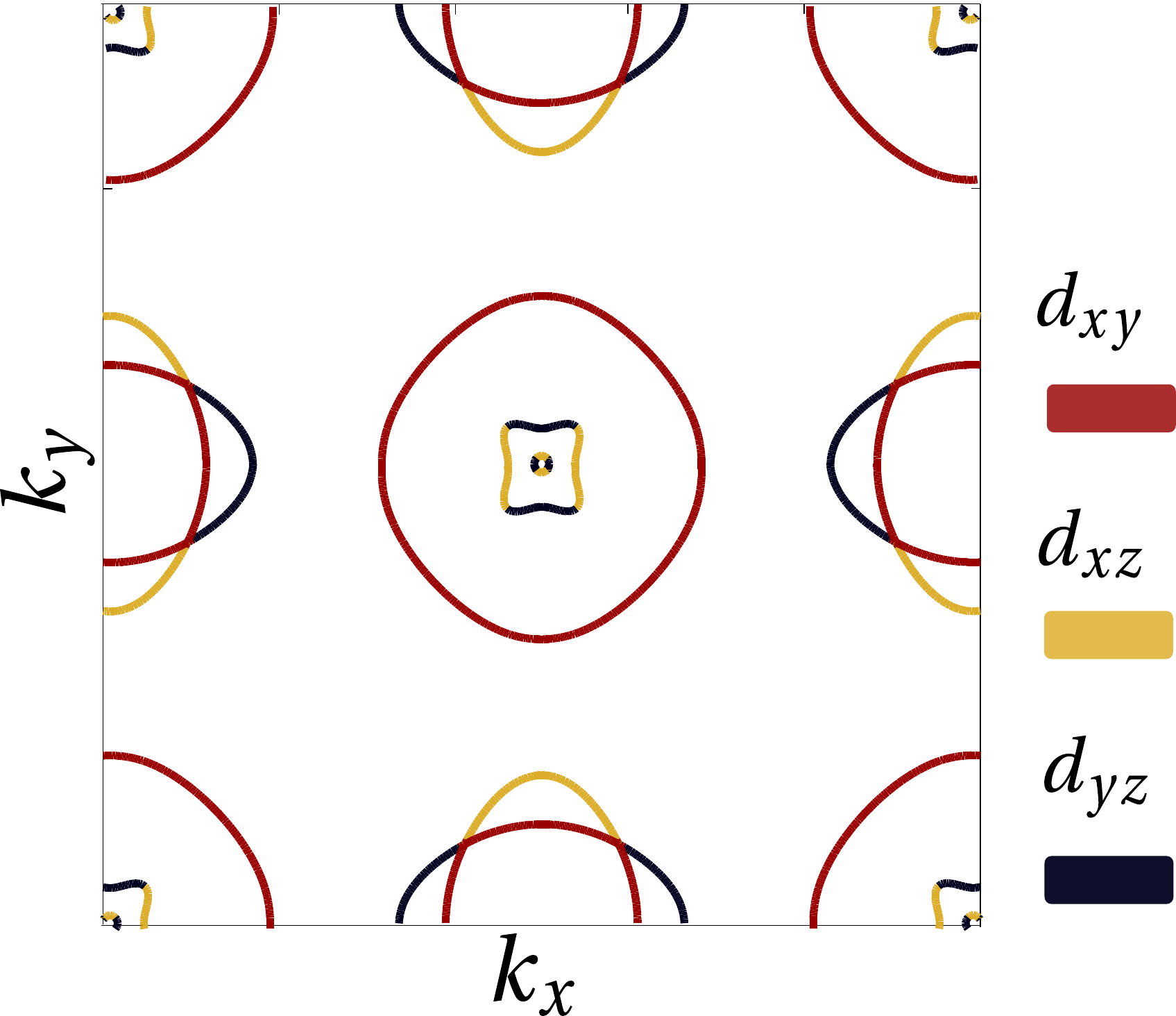}\caption{Orbitally-resolved
Fermi surface cuts for $k_{z}=\pi$. The contribution of the three dominant
orbitals $d_{xy}$, $d_{xz}$, $d_{yz}$ is shown.}%
\label{fig:orbitals}%
\end{figure}

Another very interesting proposal put forward for LiFeAs is the so-called
orbital-antiphase superconducting gap\cite{Yin2014}, originally proposed for
the iron-based superconductors within an effective model\cite{Lu2012}. In this
essentially strong-coupling approach, the phase structure of the
superconducting gap is defined not on the Fermi surfaces but on the orbitals. This
scenario appears plausible for LiFeAs due to the strong orbital
differentiation of the Fermi surface sheets. In particular, we present the
Fermi surface cuts for $k_{z}=\pi$ with orbitally-resolved matrix elements in
Fig. \ref{fig:orbitals}. For example, the $C$-state discussed in the previous
subsection can also be regarded as a type of orbital antiphase state because
the superconducting gap on the large hole pocket, which has mostly
$xy$-orbital character, has opposite sign to the other pockets, which mostly
consist of the hybridized $xz,yz$- and also $xy$-orbitals. 
Nevertheless, the $C-$state still appears within the solution of
the BCS-type of equations in the band basis, assuming the superconductivity
arises as an instability of the electron gas with respect to the actual Fermi
surface. Therefore, assuming a sign-changing but otherwise constant gap on the
orbitals must result in a rather strong angular dependence of the
superconducting gap on the Fermi surface. The question to ask then is whether such
a sign change between orbitals can be identified in  QPI experiments. To do
so, we adopted the scheme in which we introduce the superconducting gaps in
the orbital representation before the unitary transformation from the orbital
to the band space is performed. For the sake of simplicity we took the orbital
analog of the $C-$state and assumed that the sign of the gap on the
$xy$-orbital is either opposite or the same to the signs of the gaps on the
$xz,yz$ orbitals. As the other two orbitals $x^{2}-y^{2}$ a 3$z^{2}-r^{2} $ do
not contribute strongly to the states at the Fermi level, we assumed their
gaps to be small and their phase structure irrelevant. In particular, we
choose $\Delta_{xy}=3.2 \text{ meV}$ and $\Delta_{xz}=\Delta_{yz}=5 
\text{ meV}$, which produce the sizes of the superconducting gaps on the actual
Fermi surfaces sheets roughly consistent with ARPES experiments.

\begin{figure}[ptb]
\includegraphics[angle=0,width=1\linewidth]{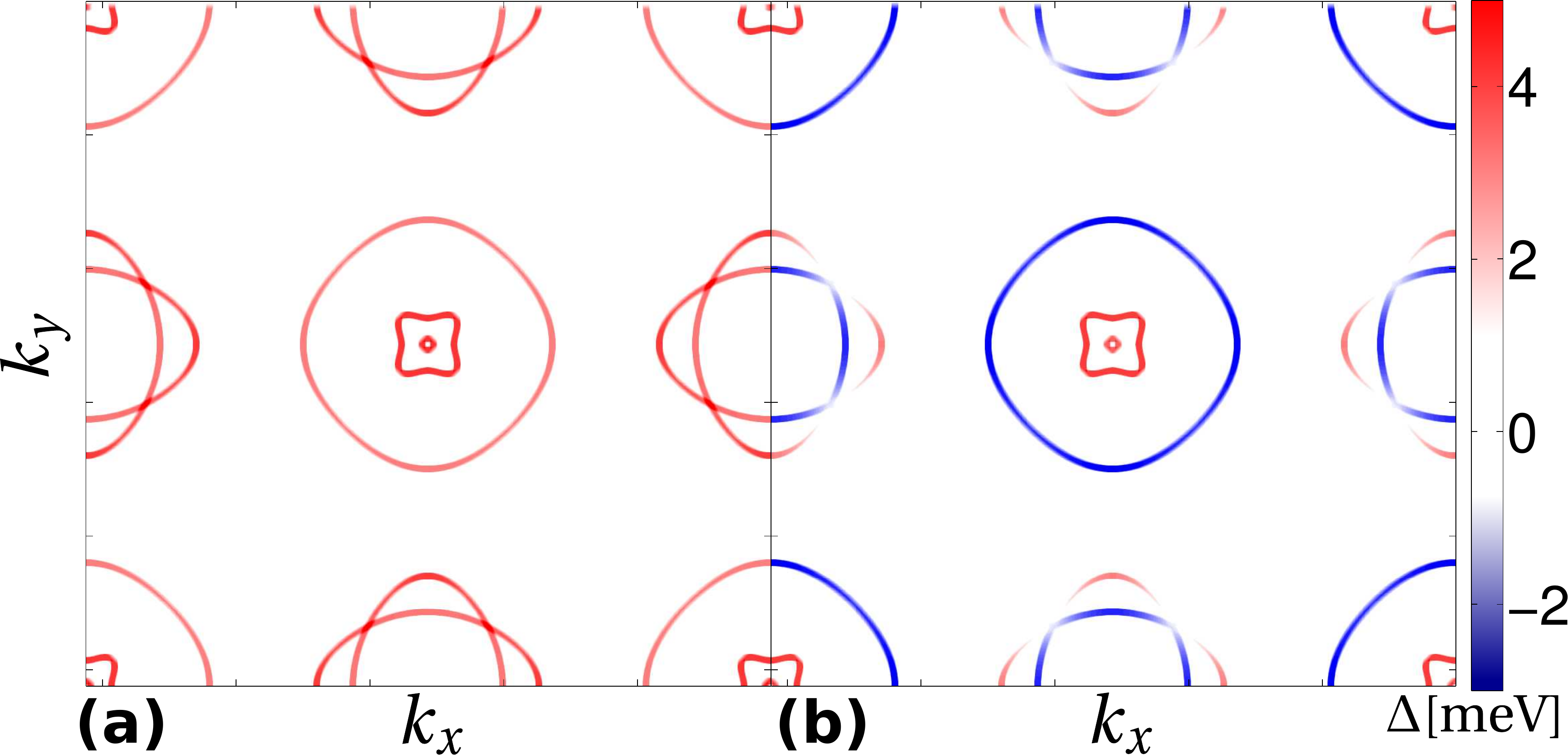}\caption{The resulting
gap structure on the Fermi surface for the $k_{z}=\pi$ cuts for the orbital
in-phase (a) and antiphase (b) superconducting gap, depicted for the LiFeAs
Fermi surface for $k_{z}=\pi$. The superconducting gaps $\Delta$ in orbital
space are projected by unitary transformation onto the band gaps on the
corresponding Fermi surface sheets. Here, the effect of the interband gaps is
neglected as it appears to be negligibly small on the Fermi surface even near
the crossing points of the two electron pockets. The resulting gap resembles
the so-called $C$-phase, except for the nodes and the sign changes in the
electron pockets.}%
\label{fig:gaponfermi}%
\end{figure}

In particular, in Fig. \ref{fig:gaponfermi} we show the resulting gap
 structure on the Fermi surface for the $k_{z}=\pi$ cuts for the
orbital in-phase (a) and antiphase (b) superconducting gap. We find clearly
anisotropic gaps on each Fermi surface sheet in both cases. Nevetherless, the
anistropy is much stronger for the $s_{+-}$-wave case. In particular, the
strongest anisotropy occurs on the electron pockets, with maximum gap values
on the tips of the elliptical pockets and deep minima or accidental nodes near
the intersection regions. Also consistent with the ARPES
experiments,\cite{sym4010251} the largest gap occurs on the tiny hole pockets.
The smallest gaps occur on the inner part of the elliptic electron
pockets and on the large hole pocket. Overall, the gap structure and its
anisotropy on the Fermi surface pockets is roughly consistent with
experimental data \cite{PhysRevLett.108.037002,sym4010251,Wang2013,Allan563},
except for the existence of nodes or deep minima near the intersections of the
two electron pockets.

In a microscopic spin fluctuation theory assuming an LDA+DMFT electronic
structure and self-energy, a pairing interaction very close to the
phenomenological orbital-diagonal pairing vertex assumed here was
calculated.\cite{Yin2014} However in that work the Fermi surface did not agree
with ARPES, and although the electron pockets were found to be fully gapped,
they were still considerably more anisotropic than found in experiment.
Furthermore, the electronic structure used in that work was not consistent
with the glide plane symmetry of the LiFeAs crystal; when some of the same
authors redid the calculation incorporating this symmetry\cite{Nourafkan2017},
they found sizable interorbital pairing matrix elements, and a conventional
$B$-type ground state as in Ref. \onlinecite{Wang2013}.
%,

\begin{figure}[ptbh]
\includegraphics[width=1\linewidth]{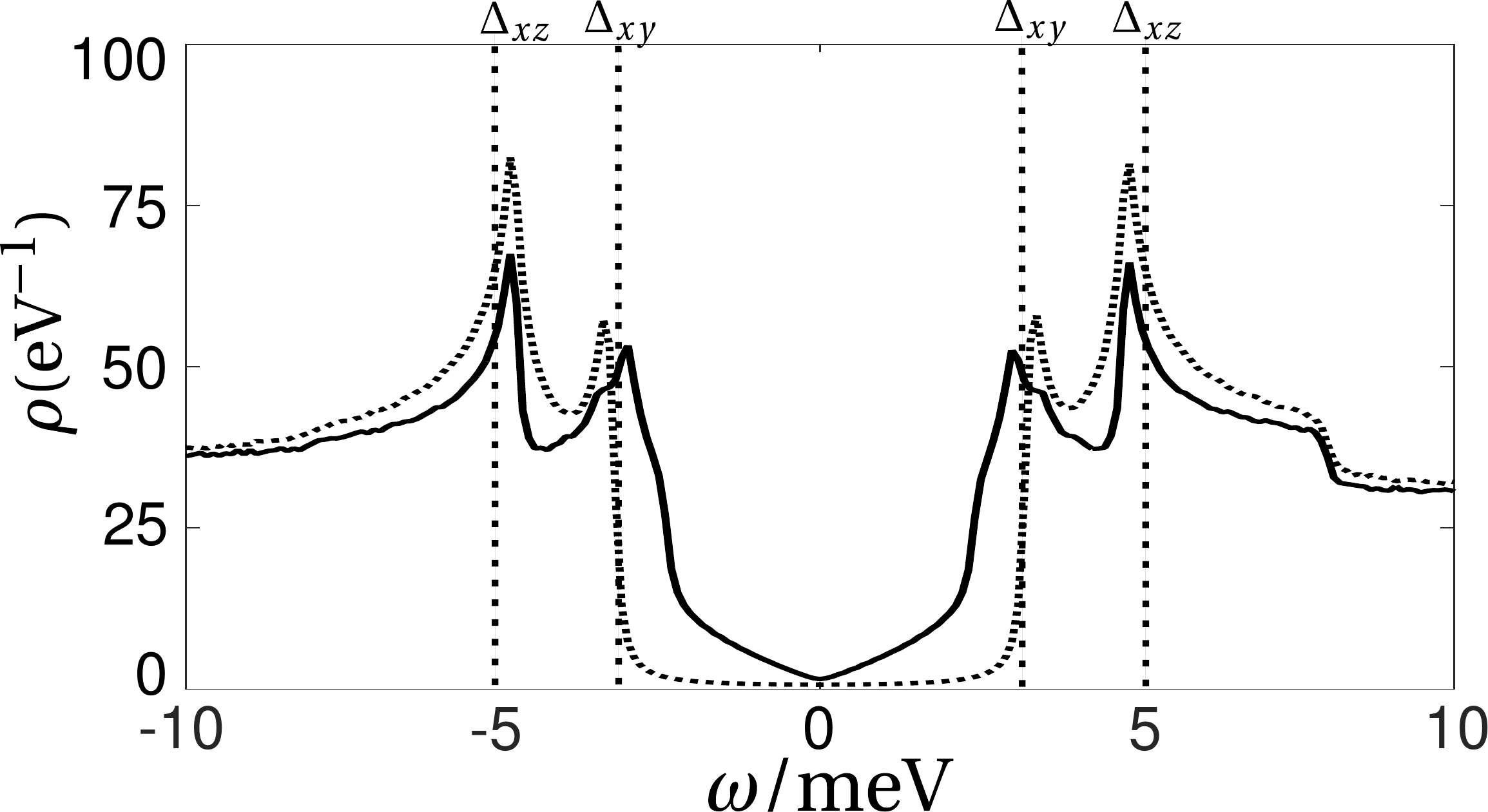}\caption{Calculated density of
states for the orbital-anti-phase ($s_{+-}$-solid curve) and orbital in-phase
($s_{++}$-dashed curve) superconducting gap, $\rho(\omega)$, using gap values
$\Delta_{xy}=3.2 \text{ meV}$, $\Delta_{xz}=\Delta_{yz}=5 \text{ meV}$.}%
\label{fig:DOSorb}%
\end{figure}

\begin{figure}[ptb]
\renewcommand{\baselinestretch}{.6}
\includegraphics[angle=0,width=1\linewidth]{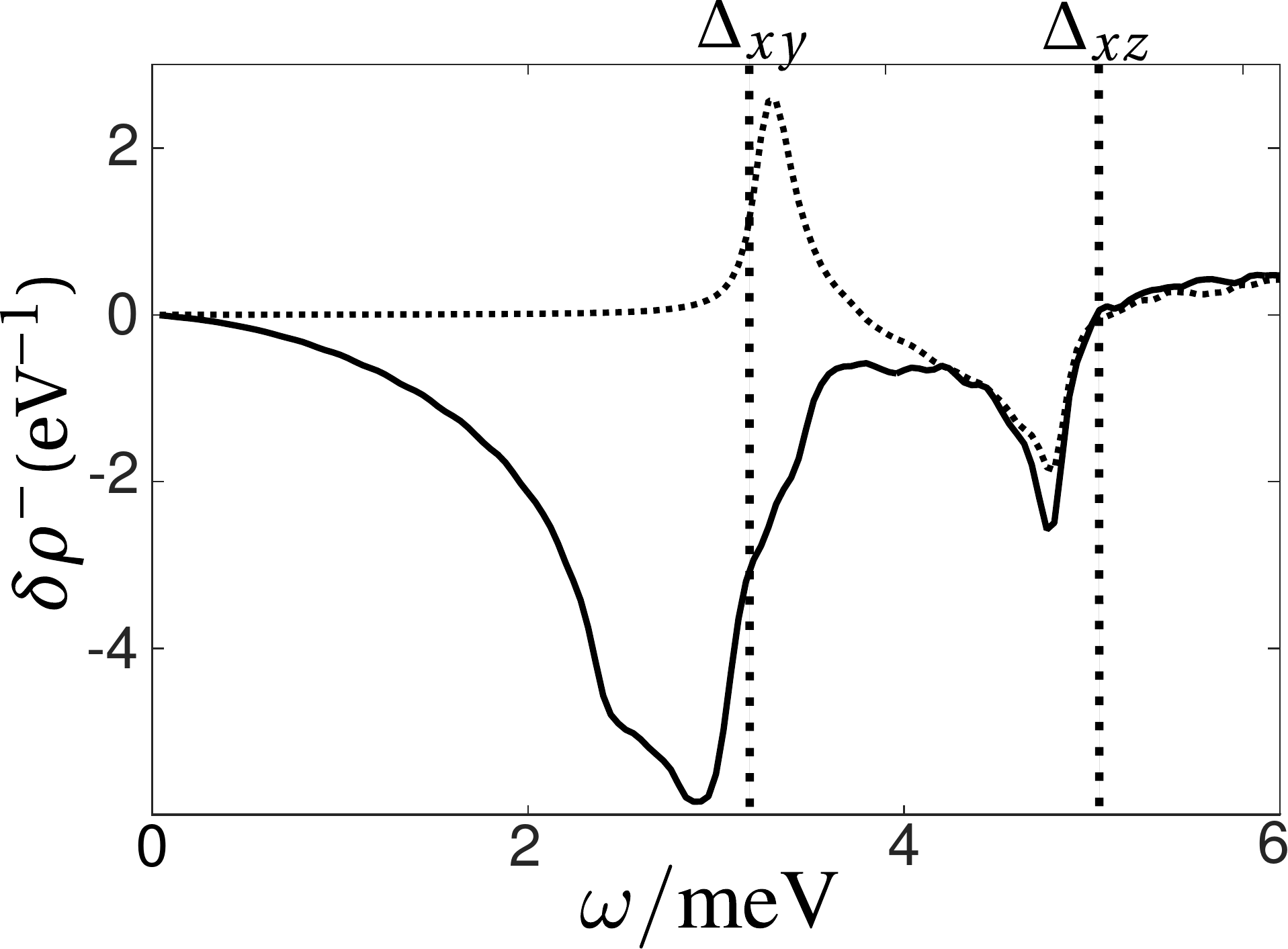}\caption{Momentum
is integrated bias-dependence of the antisymmetrized correction to the LDOS,
$\delta\rho^{-} (\omega)$ for LiFeAs at $k_{z}=\pi$. Here, we use $\Delta
_{xy}=\pm3.2 \text{ meV}$ and $\Delta_{xz}=\Delta_{yz}=5  \text{ meV}$. The sold
curve refers to the orbital antiphase $s_{+-}$, while the dashed curve is for
orbital-in-phase gap, $s_{++}$. }%
\label{fig:Fullqorbialanti}%
\end{figure}

In Fig.\ref{fig:DOSorb}, we also show the calculated density of
states for the orbital antiphase and in-phase order parameter, respectively.
Observe that the angular dependence of the gap magnitudes on the Fermi surface
is reflected in the low-energy behavior of the density of states. While the
$s_{++}$ orbital in-phase gap shows nodeless behavior at low energies, the
orbital antiphase $s_{+-}$ results in a near-nodal power-law behavior of the
density of states. 
Next we study this more complex gap structure in terms of the
correction to the LDOS due to a non-magnetic impurity diagonal in orbital
space. As we mentioned earlier, the determination of the phase structure of
the superconducting gap on the Fermi surface appears to be a non-trivial task
due to both non-zero intra- and interband gaps and strong orbital mixing on
some Fermi surface sheets. Therefore, to determine whether or not the gap
changes sign on various orbitals and not necessarily between the bands is a
challenging task. We show in Fig.\ref{fig:Fullqorbialanti} the momentum
integrated antisymmetrzed correction to the density of states, $\delta\rho^{-}
(\omega)$, in the intermediate impurity scattering limit, $U=10$ meV, for the
orbital in-phase $s_{++}$-state, and orbital antiphase $s_{+-}$-state,
respectively. Not surprisingly, we still find a clear signature of the
sign-changing gap despite the fact that the gaps reside on the orbitals rather
than on bands, but their phase structure, especially on the electron pockets,
cannot be clearly defined. Nevertheless, the behavior of the $\delta\rho
^{-}(\omega)$ within the energy region of interest (\textit{i.e.} between 3
and 5 meV) is still very characteristic.  
At the same time, the position of the peaks is less informative in
this case as it refers to the maximum of the gaps on the bands. As a result,
using QPI in this case again only allows one to determine the presence of the
sign change in the gap structure, but not the precise distribution of it among
the various bands. 

\section{Conclusion}

We have investigated an extension of the phase-sensitive method
of analyzing QPI data from non-magnetic impurities, proposed by us previously
\cite{QPI2015} in the context of a simplified model, for a realistic
10-orbital tight-binding Hamiltonian and $t$-matrix approximation for the
impurity potential. We have concentrated on studying the LiFeAs compound due
to a rather detailed experimental knowledge of the electronic structure of
this system\cite{Wang2013} and the measured superconducting gap values for
each of the Fermi surface sheets\cite{sym4010251,PhysRevLett.108.037002}. In
particular, we have shown that despite the complex Fermi surface topology, the
Friedel oscillations originating from various intraband and interband
impurity-\'{\i}nduced scattering can be clearly separated in LiFeAs.
Furthermore, the scattering within hole or electron pockets as well as between
the electron and hole Fermi surface pockets show some characteristic features,
which should help to identify them experimentally. Furthermore, the behavior
of the phase-sensitive antisymmetrized correction to the local density of
states, $\delta\rho(\omega)$, in the superconducting state allows one to
discriminate between the sign-changing and sign-preserving superconducting
gaps independent of the structure and the strength of the impurity potential.
In particular, we showed that the relative phase difference of the gap between
large hole pocket and electron pockets can be easily addressed within QPI experiments.

At the same time, the determination of the relative phase of the gap on the
tiny $Z$-centered hole pockets with respect to the other electron and hole
bands appears to be a much more subtle issue. It is still possible to
distinguish between a conventional constant sign gap and the usual $s_{+-}$
scenario where the gap is one sign on all hole pockets and another on the
electron pockets. However, once the sign structure of the gap is more complex
and involve further sign changes between large hole and small hole pockets,
the behavior of $\delta\rho^{-}(\mathbf{q},\omega)$ cannot be clearly
assigned, primarily due to the near proximity of two gap energies in this
system. We discussed the same point for the so-called orbital-antiphase
superconducting gap structure, where the gap is diagonal in orbital space,
raised in the context of the LiFeAs system\cite{Yin2014}.

In summary, the method we proposed seems to allow one to answer the question
whether the gap has a sign change, but cannot be used, at least in LiFeAs, for
a more detailed determination of the phase structure of the order parameter.

\vskip.2cm   \textit{Note added. }In the final
stages of writing, two papers appeared that discussed detailed analysis of
quasiparticle interference measurements on LiFeAs, including a new qualitative
phase-sensitive method based on interference of impurity bound
states\cite{Bonn:2017i,Bonn:2017ii}. In Ref. \onlinecite{Bonn:2017ii}, the
authors explicitly perform an analysis of the antisymmetrized $\mathbf{q}%
$-integrated differential conductance according to the prescription presented
in Ref. \onlinecite{QPI2015} and here. While they point out a substantial
quantitative dependence on the area in $\mathbf{q}$ space over which one
chooses to integrate, the qualitative result is of the \textquotedblleft
even\textquotedblright\ type between the two gap energies, in the nomenclature
of the current paper. Thus their results using our method are entirely
consistent with the results they obtain using their bound state approach, and
with the claim already made in Ref. \onlinecite{QPI2015}, namely that there is
a sign change between hole and electron pockets in LiFeAs.
\vskip .2cm
{\it Acknowledgements.}  PJH was supported by NSF-DMR-1407502, IIM by ONR through the NRL basic research program.
D.A. and I.E. were supported by the joint DFG-ANR Project (ER 463/8-1) and
DAAD PPP USA N57316180.

\bibliographystyle{apsrev4-1}
\bibliography{paper_111qpi_final}

\end{document}